\documentclass[twocolumn,prl,aps,epsfig,amsmath,superscriptaddress]{revtex4}

\usepackage{times}

\usepackage{amsmath}
\usepackage{amssymb}\usepackage{graphicx}
\usepackage[usenames, dvipsnames]{color} 
\usepackage{bm} 
\usepackage{subfigure}
\usepackage{dcolumn}
\usepackage{bm}
\usepackage{color}

\begin{document}

\title{Superlattice-induced ferroelectricity in charge-ordered La$_{1/3}$Sr$_{2/3}$FeO$_{3}$} 
\author{Se Young Park}
\affiliation{Department of Physics, University of California, Berkeley, CA 94720, USA}
\affiliation{Center for Correlated Electron Systems, Institute for Basic Science (IBS), Seoul 08826, Republic of Korea}
\affiliation{Department of Physics and Astronomy, Seoul National University (SNU), Seoul 08826, Republic of Korea}
\author{Karin M. Rabe}
\email{kmrabe@physics.rutgers.edu}
\affiliation{Department of Physics \& Astronomy, Rutgers University, Piscataway, NJ 08854, USA}
\author{Jeffrey B. Neaton}
\affiliation{Department of Physics, University of California, Berkeley, CA 94720, USA}
\affiliation{Molecular Foundry, Lawrence Berkeley National Laboratory, Berkeley, CA 94720, USA}
\affiliation{Kavli Energy NanoSciences Institute at Berkeley, Berkeley, CA 94720, USA}

\begin{abstract}
Charge-order-driven ferroelectrics are an emerging class of functional materials, distinct from conventional ferroelectrics, where electron-dominated switching can occur at high frequency. Despite their promise, only a few systems exhibiting this behavior have been experimentally realized thus far, motivating the need for new materials. Here, we use density functional theory to study the effect of artificial structuring on mixed-valence solid-solution La$_{1/3}$Sr$_{2/3}$FeO$_{3}$ (LSFO), a system well-studied experimentally. Our calculations show that A-site cation (111)-layered LSFO exhibits a ferroelectric charge-ordered phase in which inversion symmetry is broken by changing the registry of the charge order with respect to the superlattice layering. The phase is energetically degenerate with a ground-state centrosymmetric phase, and the computed switching polarization is 39 $\mu$C/cm$^{2}$, a significant value arising from electron transfer between Fe ions. Our calculations reveal that artificial structuring of LSFO and other mixed valence oxides with robust charge ordering in the solid solution phase can lead to charge-order-induced ferroelectricity.
\end{abstract}

\pacs{73.21.Cd, 75.25.Dk, 77.55.Px} 

\maketitle

The formulation of new design principles for ferroelectric materials has recently attracted great interest. A broad principle, proposed by Khomskii and co-workers \cite{Efremov04,Khomskii06}, is to start with a high-symmetry reference phase and combine two symmetry-breaking orderings, neither of which separately lift inversion symmetry, to generate two or more polar variants. Perovskite oxides are ideal systems to search for realizations of mechanisms governed by the above-mentioned broad principle. They exhibit various symmetry-lowering lattice instabilities as well as magnetic, charge and orbital ordering in bulk phases \cite{Imada98,Tokura00,Khomskii14}. Furthermore, with the recent progress in atomic-scale layer-by-layer growth techniques, symmetry-breaking compositional order can be achieved in a wide variety of complex oxide systems via superlattices \cite{Zubko11,Hwang12,Chakhalian14,Young16}. The combination of orderings has thus been the basis of discovery of several new types of perovskite ferroelectrics. In particular, in hybrid improper ferroelectricity \cite{Benedek11}, a lattice distortion that preserves inversion symmetry (typically an oxygen-octahedron rotation-tilt pattern) combines with symmetry breaking by layering, either in a Ruddlesden-Popper phase \cite{Benedek11} or in a perovskite ABO$_{3}$/A$^\prime$BO$_{3}$ (001) superlattice \cite{Bousquet08,Benedek12,Varignon15}, generating polar variants. In these systems, the switching polarization is generated by a polar lattice distortion in the lowered-symmetry state.

Charge-order-driven ferroelectricity can be obtained by combining symmetry breaking by charge ordering with symmetry breaking by layered cation ordering to generate polar variants \cite{Yamauchi09,Kobayashi12,Park17}. For example, in the 1:1 superlattice LaVO$_{3}$/SrVO$_{3}$, layered charge ordering of V$^{3+}$ and V$^{4+}$ combines with the layered ordering of La and Sr to break up-down symmetry normal to the layers, generating two polar variants \cite{Park17}. The distinctive characteristic differentiating charge-order-driven ferroelectricity from the displacive type is that the switching polarization arises primarily from interionic transfer of electrons when the charge ordering pattern is switched. This is accompanied by a small polar lattice distortion, which can be used as a proxy to signal the polar nature of the phase. Such electronic ferroelectrics might be useful for high-frequency switching devices given that the polarization switching timescale is not limited by phonon frequency \cite{Ishihara10,Yamauchi14}.

To promote experimental observation of switchable polarization due to charge-order-driven ferroelectricity,  the system should exhibit a strong tendency to charge disproportionation. Further,  there should be a strong tendency to charge ordering, which could be manifested as robust charge ordering already in the solid solution phase.  The iron oxide family is a class of materials that satisfies these conditions. Unlike La$_{1/2}$Sr$_{1/2}$VO$_{3}$, in which no charge ordering is observed in the solid solution down to low temperatures \cite{Inaba95}, robust charge orderings are observed in many iron oxides such as magnetite \cite{Verwey39}, hexagonal ferrite LuFe$_{2}$O$_{4}$ \cite{Ikeda05}, and perovskite CaFeO$_{3}$ \cite{Takano77} and La$_{1/3}$Sr$_{2/3}$FeO$_{3}$ solid solution \cite{Battle90}.

\begin{figure}[htbp]
\begin{center}
\includegraphics[width=1\columnwidth, angle=-0]{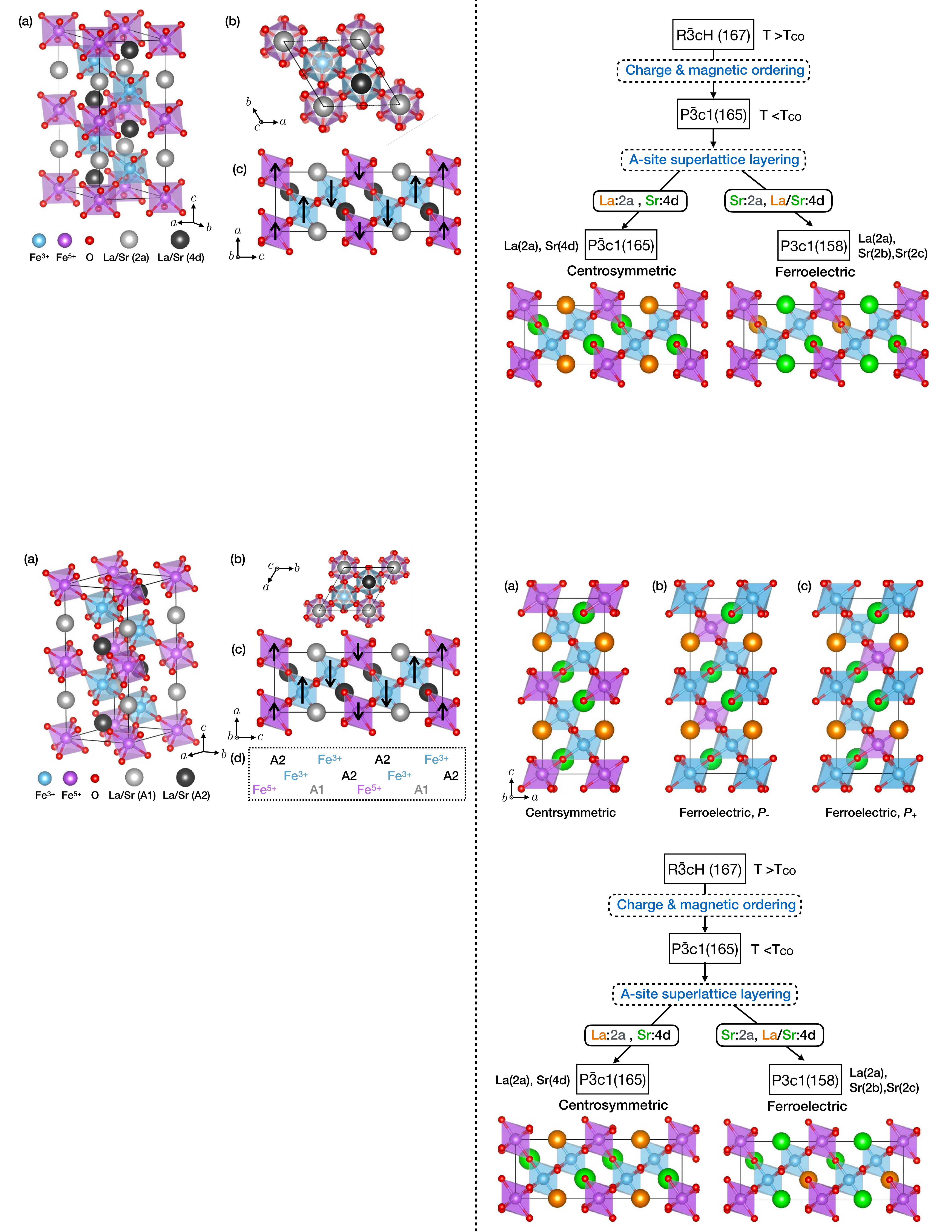}
\caption{(a) Atomic arrangement of La$_{1/3}$Sr$_{2/3}$FeO$_{3}$ solid solution with charge and magnetic ordering. Light blue and magenta colors are for Fe ions and surrounding oxygen octahedron of Fe$^{3+}$ and Fe$^{5+}$, respectively. The silver and black spheres represent the La or Sr atoms occupying two distinct Wyckoff positions ($2a$ and $4d$) of $P\bar{3}c1$ space group with multiplicity of two and four, respectively.   (b) Top view showing the $a^{-}a^{-}a^{-}$ oxygen-octahedron rotation pattern. (c) Side view showing the anti-phase antiferromagnetic ordering.}
\label{fig:str}
\end{center}
\end{figure}

Perovskite solid solution La$_{1/3}$Sr$_{2/3}$FeO$_{3}$ (LSFO) has a charge-ordered Mott-insulating state as the low temperature phase.  The average valence of Fe is +3.67 assuming +2, +3, and -2 charge states for Sr, La and O, respectively. For T $>$ 210K, LSFO is metallic with all Fe sites equivalent. Below 210K, a metal-insulator transition is observed with the onset of both antiferromagnetic (AF) and charge ordering (CO) where two distinct charge states are stabilized by the breathing distortion of the oxygen octahedra \cite{Battle90,Li97,McQueeney07}. Nominally this charge ordering corresponds to 3Fe$^{3.67+}$ = Fe$^{5+}$ + 2Fe$^{3+}$, but due to the strong hybridization of Fe-$d$ $e_g$ states and the surrounding oxygen ligands, the configurations include some ligand hole character \cite{Matsuno99} and indeed the measured magnetic moments of Fe$^{3+}$/Fe$^{5+}$ states are lower than the nominal values \cite{Battle90,McQueeney07}; however, for convenience we will continue to refer to these charge states as Fe$^{3+}$ and Fe$^{5+}$. Fig.~\ref{fig:str}(a) describes the crystal structure with the observed charge ordering pattern, in which each charge state forms a (111) plane stacked in the repeated pattern of Fe$^{5+}$-Fe$^{3+}$-Fe$^{3+}$. The magnetic moments order ferromagnetically within each (111) plane. In the out-of-plane direction, the moments order in an anti-phase antiferromagnetic (APAF) pattern in which the moments in adjacent planes with the same charge state are antialigned and the moments in adjacent planes with different charge states are aligned, as illustrated in Fig.~\ref{fig:str} (c). Experimentally, (001) and (111) oriented thin films and (001) oriented superlattices are reported, maintaining the  charge-ordered insulating phase of the bulk\cite{Minohara16,Krick16}.

Charge ordering in LSFO has been the subject of investigation through first-principles calculations \cite{Matsuno99,Saha-Dasgupta05,Krick16, Zhu2018}, with disproportionation to Fe$^{3+}$/Fe$^{5+}$ observed when on-site electron correlation is included via a Hubbard $U$ parameter \cite{Saha-Dasgupta05}. These studies have considered various ordering patterns for La/Sr and their effects on charge disproportionation, charge ordering and magnetic ordering of the electronic ground state \cite{Saha-Dasgupta05, Krick16, Zhu2018}. The energetics of the system are found to be driven by magnetic exchange, with both lattice distortion and correlation needed to produce an insulating ground state. 

In this paper, using symmetry analysis and first-principles density functional theory results, we find that a layered superlattice ordering of La/Sr cations in charge-ordered La$_{1/3}$Sr$_{2/3}$FeO$_{3}$ generates electronic ferroelectricity via the combined symmetry breaking of charge ordering and cation layering. Depending on the registry between the superlattice layering and the charge ordering patterns, both centrosymmetric and ferroelectric phases can be stabilized. Electronic and structural properties of the low energy phases and details of switching between two polar branches are presented. This example serves as a proof of concept for this design principle, which can be more broadly applied in the search for electronic ferroelectricity in chemically complex perovskite oxide solid solutions with multiple valence cations.

\begin{figure}[htbp]
\begin{center}
\includegraphics[width=0.9\columnwidth, angle=-0]{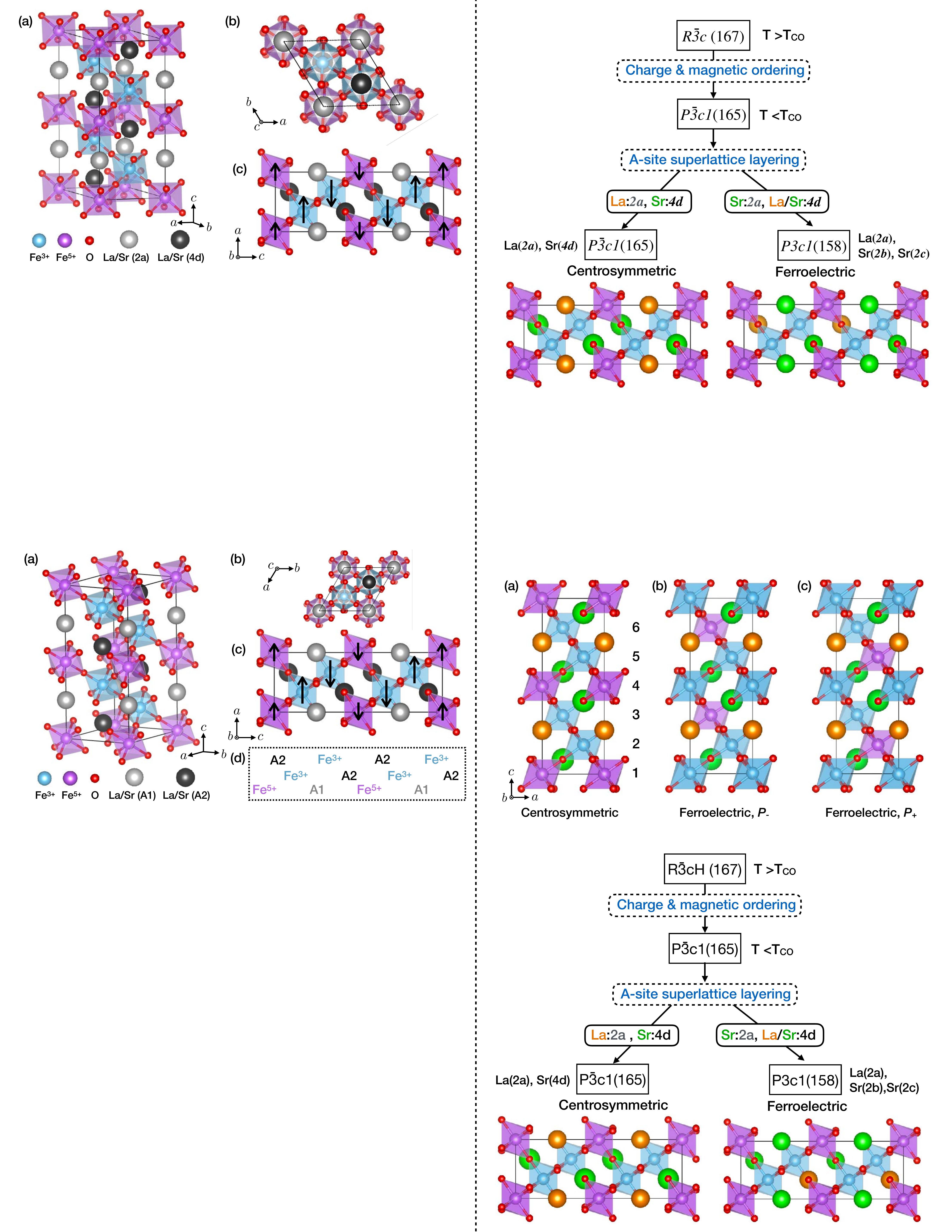}
\caption{Space group analysis of symmetry lowering in the LSFO solid solution. Starting from the high-temperature $R\bar{3}c$ structure, the charge ordering lowers the symmetry to $P\bar{3}c1$ space group. From there, different superlattice layering arrangements described in the text produce centrosymmetric and ferroelectric phases.}
\label{fig:space-group}
\end{center}
\end{figure}

We perform first-principles density-functional-theory calculations with the generalized gradient approximation plus $U$ (GGA+$U$) method using the Vienna {\it ab-initio} simulation package \cite{Kresse96,Kresse99}. The Perdew-Becke-Erzenhof parametrization \cite{Perdew96} for the exchange-correlation functional and the rotationally invariant form of the on-site Coulomb interaction \cite{Liechtenstein95} are used with $U=5.4$ eV and $J=1.0$ eV for the iron $d$ consistent to the values used  ($U_{eff} (=U-J)$= 4.4 eV \cite{Krick16}, $U=$ 5 eV, $J=$ 1 eV \cite{Saha-Dasgupta05,Zhu2018} ) ; and $U_{f}=11$ eV and $J_{f}=0.68$ eV are used for the La $f$ states to shift the La $f$ bands away from the Fermi energy \cite{Czyzyk94}. The charge ordering pattern is determined through Wannier center analysis. We find that the stability of Fe$^{3+}$/Fe$^{5+}$ charge ordering with APAF magnetic ordering is insensitive to the broad range of the $U$ values (see the Supplementary Information). We used the projector augmented wave method \cite{Blochl94} with pseudopotential containing 6 valence electrons for O $(2s^{2}2p^{4})$, 14 for Fe $(3p^{6}3d^{7}4s^{1})$, 11 for La $(5s^{2}5p^{6}5d^{1}6s^{2})$, and 10 for Sr $(4s^{2}4d^{6}5s^{2})$ in which the same pseudopotential are used for Fe ions regardless of charge states. For each charge ordering pattern, we construct a starting crystal structure with breathing distortions corresponding to the charge-ordering pattern (see the Supplementary Information for for further details). A simple $\sqrt{2} \times \sqrt{2} \times 2\sqrt{3}$ hexagonal unit cell with six Fe atoms per cell is chosen to accommodate the relevant octahedral rotation distortions and charge-order patterns. An energy cutoff of 600 eV, $k$-point sampling on a $\Gamma$-centered $6\times 6\times 3$ grid and a force threshold of 0.01 eV/\AA are used for full structural relaxation. The ferroelectric polarization is calculated using the Berry phase method \cite{King-Smith93} with $8\times 8\times 4$ $k$-point grid.

\begin{table*}[htbp]
\caption{\label{tab:en-vs-moco} Low energy phases of LSFO (111) superlattice. For each phase, the charge ordering, total energy per Fe, electronic character, magnetic ordering and magnetic moments are given.  I and M denote insulator and metal, respectively. The symbols F, AF, Fi, and APAF denote ferromagnetic, antiferromagnetic, ferrimagnetic, and antiphase-antiferromagnetic ordering, respectively. For the magnetic ordering patterns, the arrows represent spin directions and magnitude in which the double arrows denote the Fe$^{3+}$ charge state and single arrows denote the other charge states (Fe$^{5+}$ or Fe$^{4+}$). The calculated magnetic moments presented are for the first three Fe sites.}
\centering
\begin{tabular}{c|cccccccc}
Charge States &\multicolumn{6}{c}{2Fe$^{5+}(d^{3})$/4Fe$^{3+}$($d^{5}$)}  &6Fe$^{3.67+}$($d^{4.33}$)  & \multicolumn{1}{c}{2Fe$^{3+}$($d^{5}$)/4Fe$^{4+}$($d^{4}$)} \\
\hline
CO patterns  &CS & FE & CS & FE & CS  & FE & CS & CS  \\
\hline
$\Delta E/$Fe (meV)&0 & 0.54 &124 & 119 & 243 & 42 & 24  & 51 \\
Phase & I & I & I & I & I  & I & M  & M \\
Mag. ordering & APAF & APAF  & AF & AF & Fi & Fi2  &F &   Fi  \\
& $\uparrow\Uparrow\Downarrow\downarrow\Downarrow\Uparrow$ & $\Uparrow\uparrow\Uparrow\Downarrow\downarrow\Downarrow$ & $\uparrow\Downarrow\Uparrow\downarrow\Uparrow\Downarrow$& $\Uparrow\downarrow\Uparrow\Downarrow\uparrow\Downarrow$& $\uparrow\Downarrow\Downarrow\uparrow\Downarrow\Downarrow$ 
&  $\Uparrow\Downarrow\uparrow\Uparrow\Downarrow\uparrow$ &$\uparrow\uparrow\uparrow\uparrow\uparrow\uparrow$ &  $\Uparrow\downarrow\downarrow\Uparrow\downarrow\downarrow$ \\
$m$ [Fe1/Fe2/Fe3] $(\mu_{B})$ &3.5/4.0/4.0 & 4.0/3.5/4.0 & 2.8/4.1/4.1 & 4.1/2.9/4.1 & 2.7/4.2/4.2 &  3.9 / 4.1/ 3.3  & 4.0/ 4.0/ 4.0 &4.1/3.7/3.7  
\end{tabular}
\end{table*}

Our analysis of the symmetry breaking by charge ordering and cation ordering in this system starts with the atomic arrangement in the LSFO solid solution shown in Fig.~\ref{fig:str}. The corner-connected octahedra of the perovskite structure are rotated in the $a^{-}a^{-}a^{-}$ pattern with La/Sr ions in between the Fe-centered octahedral cages, forming a triangular lattice as viewed along $c$ (pseudo cubic [111]) direction (Fig.~\ref{fig:str} (b)). This high-symmetry phase has space group $R\bar{3}c$, with all Fe sites equivalent and all A cation sites equivalent.  We analyze the symmetry lowering due to the charge and cation orderings in Fig.~\ref{fig:space-group}. First, we introduce the experimentally observed (111)-oriented charge ordering, which lowers the space group symmetry to $P\bar{3}c1$ in which there are two Wyckoff positions for A-site ions ($2a$, $4d$), represented as silver and black spheres in Fig.~\ref{fig:str}(a). Next, we introduce layered A-site cation orderings. Following the (111)-oriented charge ordering, relevant A-site orderings are expected to be (111)-oriented and thus can be expressed within the 30-atom unit cell. In order to maintain the same average valence for Fe-$d$ states, the superlattice needs to satisfy the constraint of 1:2 ratio of La and Sr. This yields two types of atomic arrangements: one maintaining the same symmetry by assigning La to $2a$ (with multiplicity 2) and Sr to $4d$ (with multiplicity 4) Wyckoff positions, and the other lowering symmetry by assigning more than one atomic species to the $2a$ and/or $4d$-Wyckoff position. The first type, illustrated in the bottom left of Fig.~\ref{fig:space-group}, is clearly unique and does not break any additional symmetry. The latter type allows for many possible arrangements, but there is only one arrangement, shown in the bottom right of Fig.~\ref{fig:space-group}, that has uniformly separated La-(111) planes which, as observed for the (001)-oriented superlattice \cite{Krick16}, maintains the charge ordering pattern. For this arrangement, the space group symmetry is lowered to non-centrosymmetric $P3c1$, demonstrating the possibility of superlattice-driven ferroelectricity in the presence of charge ordering.

\begin{figure}[htbp]
\begin{center}
\includegraphics[width=1\columnwidth, angle=-0]{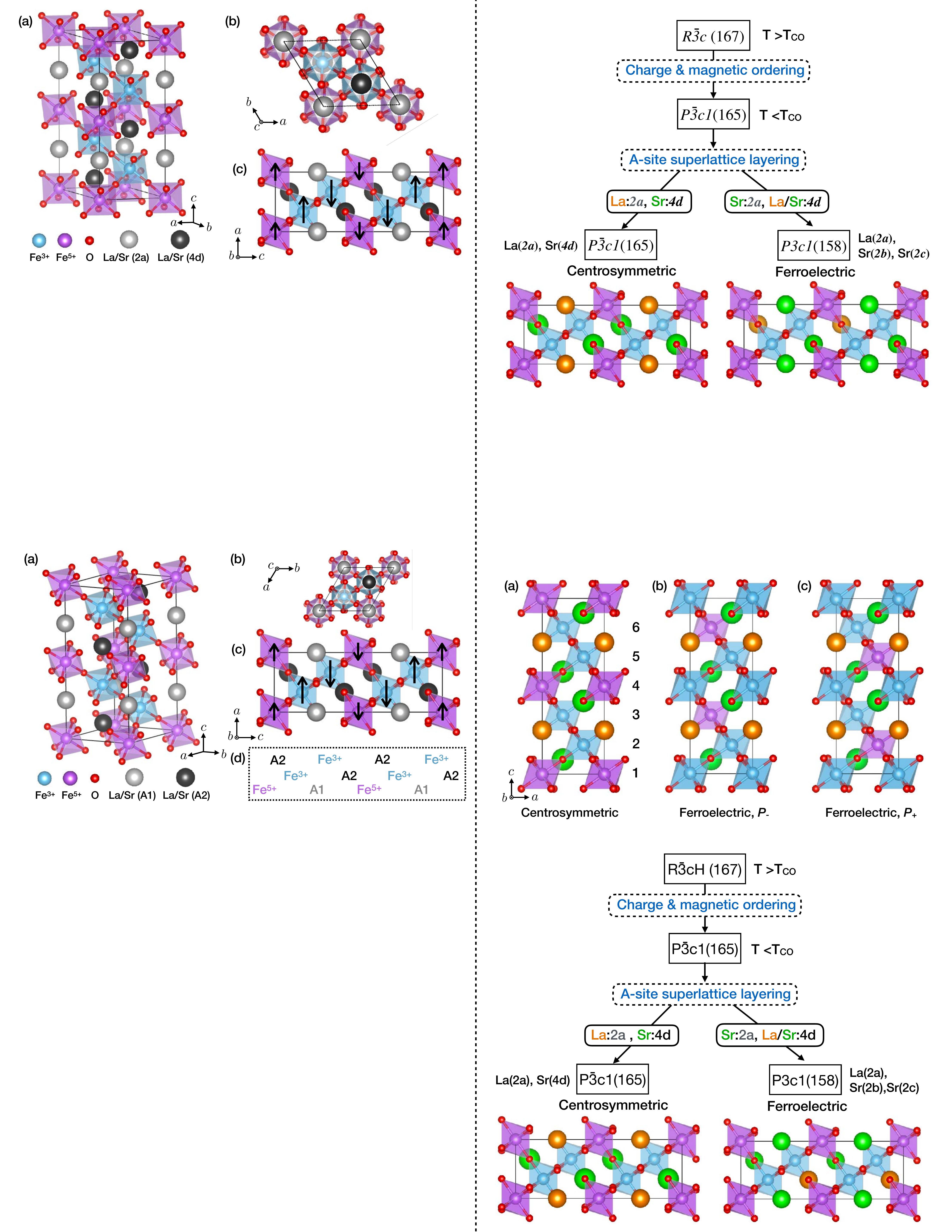}
\caption{Three charge ordering patterns for A-cation ordered superlattices. (a) Centrosymmetric charge ordering pattern. (b-c) Ferroelectric charge ordering patterns with two polar variants in which  $P_{+}/P_{-}$ are related by inversion. The Fe layers are numbered 1-6. Both for the centrosymmetric and ferroelectric phases, the octahedral rotations along the $c$-direction are opposite around the pairs of Fe sites in the layer (1,4), (2,5), and (3,6).}
\label{fig:co}
\end{center}
\end{figure}

Fig.~\ref{fig:co} shows the charge ordering patterns for the centrosymmetric $P\bar{3}c1$ phase (panel (a)) and the two polar variants of the ferroelectric $P3c1$ phase (panel (b) and (c)). We can see that the main difference between the centrosymmetric and ferroelectric phases is the registry of the charge ordering with respect to A-site layered ordering. This suggests that electric-field-controlled switching from the centrosymmetric to the ferroelectric phase or between two polar variants in the ferroelectric phase can occur through charge transfer between Fe sites. 

With first-principles calculations, we now investigate the total energies and structural parameters of these superlattice phases. In addition, we check the assumption that the charge ordering pattern of the solid solution is maintained in the A-site ordered structures. We generate the alternative charge and magnetic ordering patterns to be considered through a systematic symmetry analysis. The details of this analysis and of the first-principles approach are discussed in the Supplementary Information. 

Table \ref{tab:en-vs-moco} summarizes the low-energy phases identified. The lowest energy phase is the centrosymmetric (CS) $P\bar{3}c1$ phase (Fig.~\ref{fig:co}(a)) with Fe$^{3+}$/Fe$^{5+}$ charge ordering and APAF magnetic ordering; it is an insulator with a DFT+$U$ band gap of 0.3 eV. This charge ordering and magnetic ordering pattern is identical to that observed in the solid solution, confirming the hypothesis that the charge and magnetic ordering would be insensitive to the A-site ordering. The oxygen octahedral volume in the relaxed structure of CS-APAF phase is 9.5 \AA$^{3}$ for Fe$^{5+}$ and 10.5 \AA$^{3}$ for Fe$^{3+}$, showing the breathing distortion. The Fe$^{3+}$ ions shift about 0.01{\AA} from the center of the octahedron toward the La plane. 
The ferroelectric (FE) $P3c1$ phase (Fig.~\ref{fig:co} (b) and (c)) with the same APAF magnetic ordering is only 0.54 meV/Fe higher in energy, with octahedral distortions and magnetic moments for Fe$^{3+}$ and Fe$^{5+}$ virtually identical to those in the CS-APAF phase. With respect to the electrostatic energy, the main difference between the two different registries of the charge ordering pattern with respect to the A-site layered ordering is that in the CS phase, an Fe$^{5+}$ plane lies between the two adjacent Sr$^{2+}$ layers, while in the FE phase, the Fe$^{5+}$ plane lies between a Sr$^{2+}$ and a La$^{3+}$ layer, with a nominally higher electrostatic energy. The tiny computed energy difference implies substantial screening from the oxygen ligands, also seen in other charge-order-induced ferroelectric materials \cite{Park17}. As a result, it should be possible to switch this system to a polar variant with an applied external field with retention of the polar structure when the field is removed, producing a ferroelectric $P$-$E$ hysteresis loop.

Considering magnetic ordering patterns, for ferromagnetic ordering we find a metallic $P\bar{3}c1$ phase without charge ordering,  suggesting that insulating character is induced by the charge ordering. For antiferromagnetic and ferrimagnetic ordering, we find charge-ordered insulating phases substantially higher in energy ($>$100 meV/Fe) than the CS-APAF phase.  The energy increase on changing the magnetic ordering patterns is consistent with the Goodenough-Kanamori-Anderson rule \cite{Anderson50,Goodenough55,Kanamori59} favoring APAF magnetic ordering, and the relatively large energy increase suggests strong magnetic interactions consistent with inelastic neutron screening measurements \cite{McQueeney07}.  The presence of the charge ordering accompanied by breathing distortions for most of the magnetic phases suggests the strong electron lattice coupling driving the charge ordering, in contrast to CaFeO$_{3}$ in which the breathing distortion is strongly dependent on magnetic ordering through a proposed spin-assisted covalent bonding mechanism \cite{Cammarata12}. 

In addition to the Fe$^{5+}$/Fe$^{3+}$ charge ordering patterns, we find a phase with different pair of nominal valences Fe$^{3+}$/Fe$^{4+}$ while maintaining the metallic phase (last column of Table \ref{tab:en-vs-moco}); this phase is centrosymmetric and ferrimagnetic (CS-Fi). We note that this phase has doubly degenerate bands along high-symmetry lines crossing the Fermi energy. As discussed in more detail in the Supplementary Information, these degeneracies are protected by inversion symmetry, and breaking the inversion symmetry by changing the registry of the superlattice layering and the charge order leads to the FE-Fi2 phase, opening a band gap.

\begin{figure}[t]
\begin{center}
\includegraphics[width=0.9\columnwidth, angle=-0]{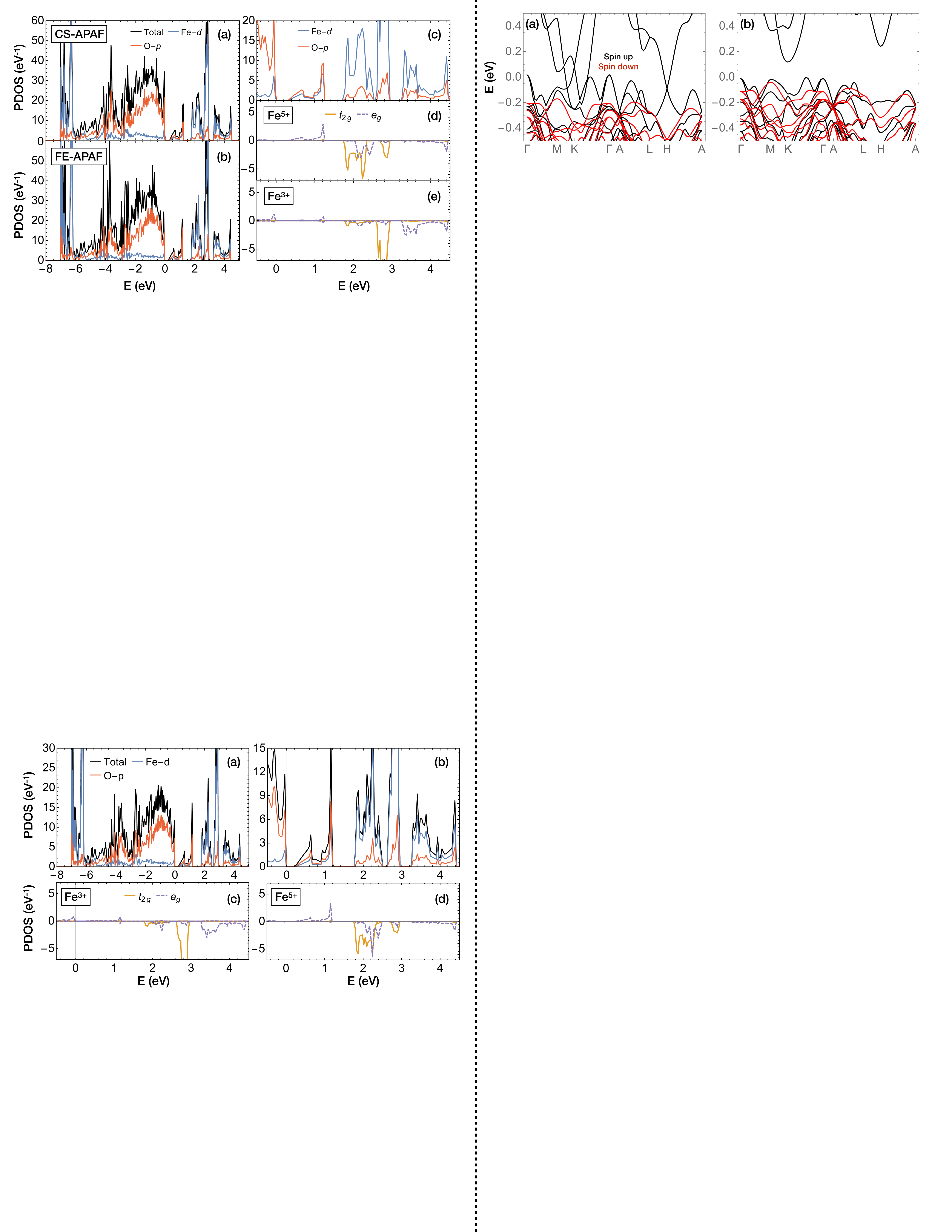}
\caption{Total density of states (black line) and orbital projected density of states (PDOS) for Fe-$d$ (blue) and O-$p$ (red) orbitals of CS-APAF (panel (a)) and FE-APAF phases (panel b).  (c) Close-up of PDOS for the unoccupied states. (d) PDOS of unoccupied $t_{2g}$ and $e_{g}$ states for spin-up Fe$^{5+}$ site. (e) PDOS of unoccupied $t_{2g}$ and $e_{g}$ states for spin-up Fe$^{3+}$ site.}
\label{fig:pdos}
\end{center}
\end{figure}

Fig.~\ref{fig:pdos}(a) shows the projected density of states (PDOS) of the CS-APAF phase; we find that the PDOS of FE-APAF phase (Fig.~\ref{fig:pdos}(b)) is almost identical to CS-APAF phase, consistent with the small energy and structural difference. The valence bands are mainly derived from O-$p$ bands located above the occupied Fe-$d$ bands around -7 eV and the conduction bands consist mostly of Fe-$d$ derived bands with a band gap about 0.3 eV, showing that this is a charge-transfer insulator \cite{Zaanen85}. We find that the low-lying unoccupied bands around 1 eV have strong hybridization between Fe-$d$ and O-$p$ states (Fig.~\ref{fig:pdos}(c)). From the PDOS of Fe$^{5+}$ (Fig.~\ref{fig:pdos} (d)), we find that the low-lying states consist of the unoccupied $e_{g}$ states hybridized with O-$p$ states, supporting the strong screening by the oxygen ligand holes, previously discussed for the solid solution \cite{Matsuno99,Yang03}. The PDOS of Fe$^{3+}$ shows negligible occupation of spin-up states, consistent with a fully spin-polarized $d^{5}$ state.

\begin{figure}[t]
\begin{center}
\includegraphics[width=1\columnwidth, angle=-0]{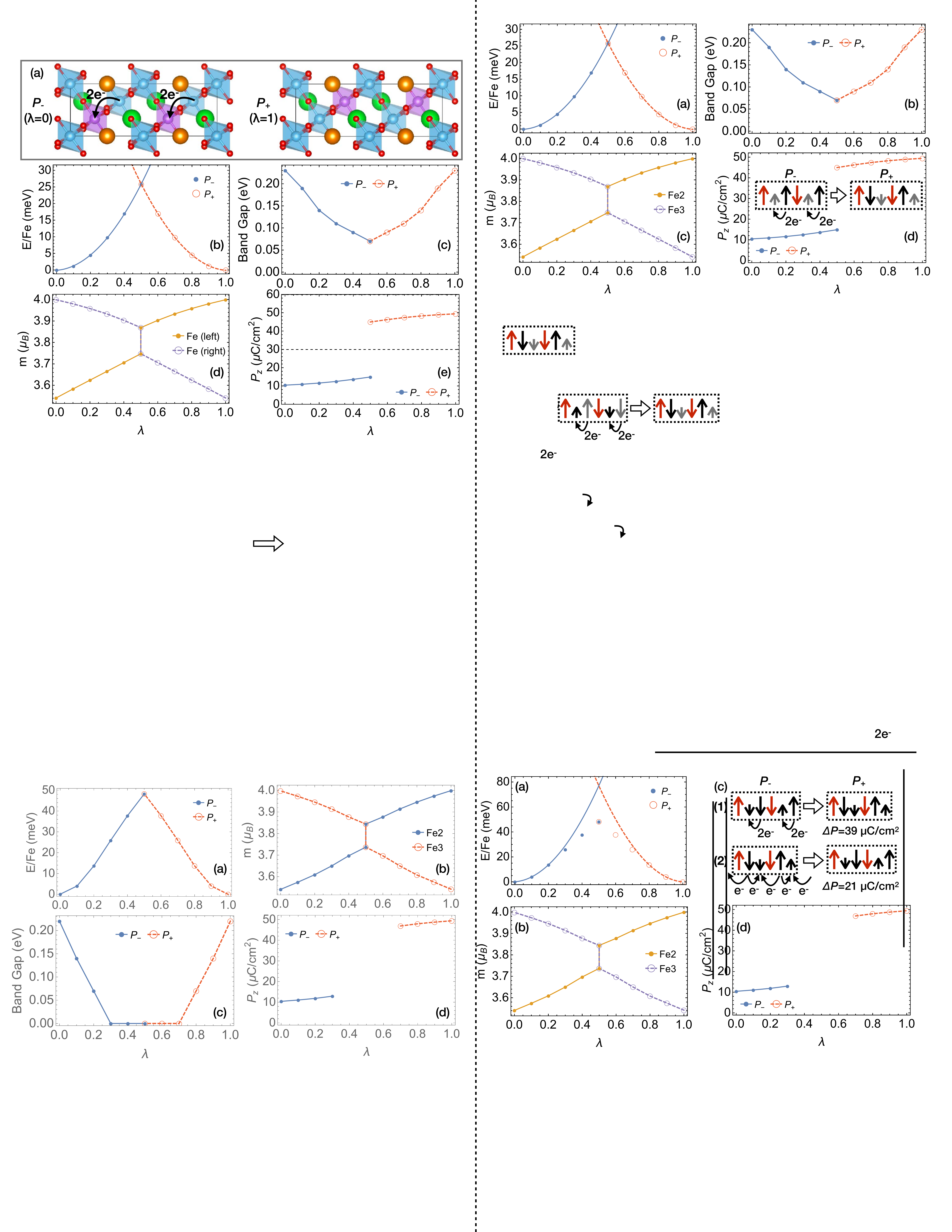}
\caption{(a) Atomic arrangements of two polar variants showing the electron transfer relating the two. (b) Total energy per Fe with respect to the lowest energy configuration for structures linearly interpolated between two inversion-related polar variants ($\lambda=0$ for $P_{-}$ and $\lambda=1$ for $P_{+}$) of the FE-APAF phase. The blue solid line and red dashed lines in panels (b), (c), and (e) denote states converged from $\lambda=0$ and $\lambda=1$ states, respectively. (c) $\lambda$ dependence of the DFT+$U$ band gap. (d)  $\lambda$ dependence of the magnetic moments of the two Fe sites adjacent to the La layer (left and right in (a)). (e) Polarization parallel to $c$-axis as a function of $\lambda$ assuming a branch choice corresponding to the electron transfer in (a). The dashed line denotes the value of half quantum of polarization.}
\label{fig:switching}
\end{center}
\end{figure}

Fig.~\ref{fig:switching} presents ferroelectric switching properties in the FE-APAF phase. We consider a switching between two inversion-related polar variants by investigating the linear interpolation of the structure between them.  The calculated energy barrier along the path is 25 meV per Fe; this is smaller than the barriers calculated for small polaron hopping for Li$_{x}$FePO$_{3}$ (88-108 meV) and hematite (85-120 meV) \cite{Maxisch06,Adelstein14}, approaching the computed double well potential depth in the conventional ferroelectric BaTiO$_3$ (20 meV) \cite{Cohen92}. The cusp in the Born-Oppenheimer energy surface indicates a first-order transition between the two polar branches, with a band gap $>$ 70 meV maintained along the transition path (Fig.~\ref{fig:switching}(c)). The first-order nature of the transition can be also seen from the magnetic moments of the Fe ions, with discontinuous jumps at $\lambda=0.5$ from the nominal 2e$^{-}$ charge transfer (Fig.~\ref{fig:switching}(d)). This coincides with the discontinuous jump in the polarization, shown in Fig.~\ref{fig:switching}(e) assuming the charge transfer of 2e$^-$ across the La layer, supported by Wannier center interpretation of the switching polarization in which the number of Wannier centers of Fe-$d$ derived occupied bands change by 2 at $\lambda=0.5$ (for further discussion see Supplementary Information). The magnitude of the switching polarization is 39 $\mu$C/cm$^{2}$, comparable to 2$P_s$ = 54 $\mu$C/cm$^{2}$ of BaTiO$_3$.

In conclusion, we propose a design principle for charge-order induced ferroelectricity by artificial structuring of mixed valence oxides with robust charge ordering in the solid solution phase. We have demonstrated this principle for charge-ordered Mott insulating La$_{1/3}$Sr$_{2/3}$FeO$_{3}$ solid solution in which a substantial switching polarization is predicted due to charge transfer between Fe sites. Our approach is not limited to perovskite oxides and can be applied to broad classes of mixed valence materials to identify new electronic ferroelectrics, potentially useful for applications requiring high-frequency switching properties.

\begin{acknowledgments}
We thank J. Ahn, C. Dreyer, A. Georges, D. Khomskii, H. -S. Kim, J. -H. Lee, K. Lee, S. May, A. J. Millis, T. W. Noh, J. -M. Triscone, D. Vanderbilt, and B. -J. Yang, for valuable discussion. This work is supported by the Materials Project funded by the U.S. Department of Energy, Office of Science, Office of Basic Energy Sciences, Materials Sciences and Engineering Division under Contract No. DE-AC02-05-CH11231: Materials Project program KC23MP. This work is also supported by the Institute for Basic Science in Korea (Grant No. IBS-R009-D1), Office of Naval Research grant N00014-17-1-2770. This work is also supported by the Molecular Foundry through the US Department of Energy, Office of Basic Energy Sciences, under the same contract number. Computational resources were provided by DOE (LBNL Lawrencium) and the Rutgers University Parallel Computer (RUPC) cluster.
\end{acknowledgments}

\bibliography{LSFO.bib}

\pagebreak

\onecolumngrid
\newpage

\begin{center}
\textbf{\large Supplementary Information: Superlattice-induced ferroelectricity in charge-ordered La$_{1/3}$Sr$_{2/3}$FeO$_{3}$}
\end{center}
\setcounter{equation}{0}
\setcounter{figure}{0}
\setcounter{table}{0}
\setcounter{page}{1}
\makeatletter
\renewcommand{\theequation}{S\arabic{equation}}
\renewcommand{\thefigure}{S\arabic{figure}}
\renewcommand{\bibnumfmt}[1]{[S#1]}
\renewcommand{\citenumfont}[1]{S#1}

\section{Enumeration of charge and magnetic orderings with layered A-site ordering}

\begin{figure}[htbp]
\centering
\includegraphics[width=0.8\textwidth]{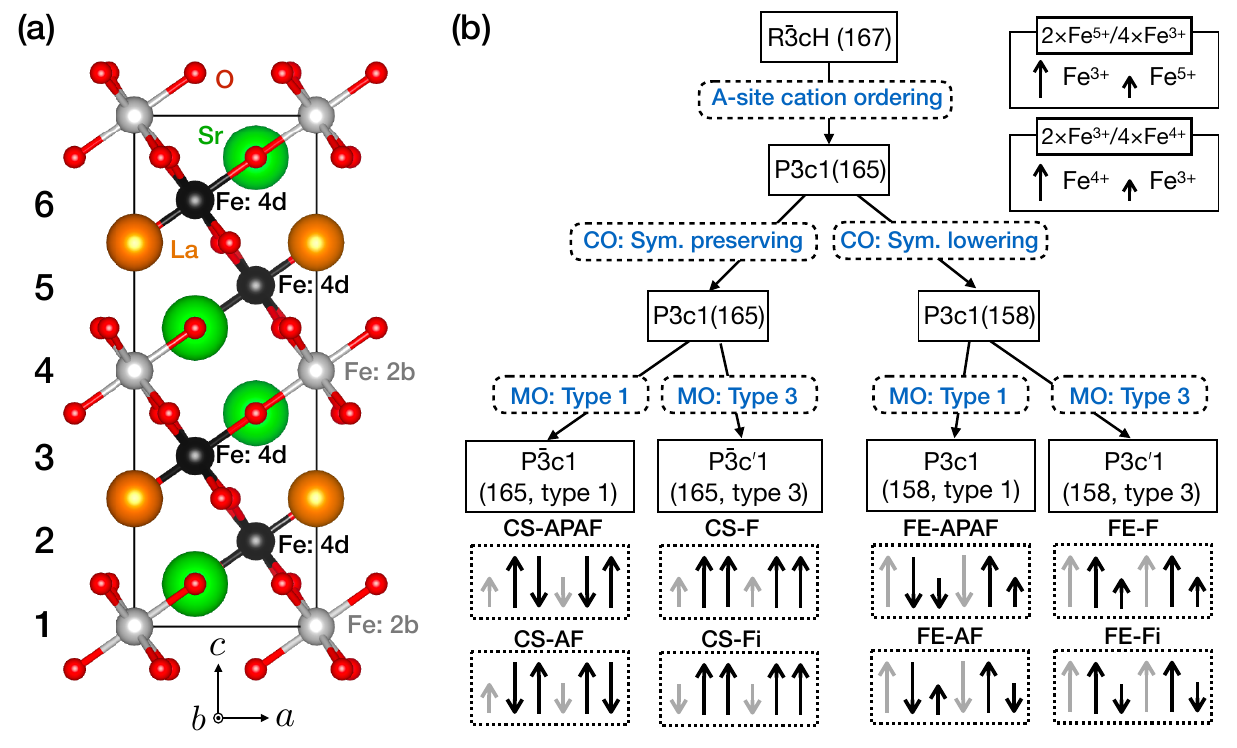}
\caption{(a) Atomic arrangement of the $P\bar{3}c1$ (165) structure of LSFO with uniformly separated La-(111) layers showing two distinct Wyckoff positions for Fe atoms, represented by silver ($2a$) and black ($4b$) spheres, with layer numbers n for Fe(n) given on the left. (b) Symmetry analysis of charge and magnetic orderings in the presence of the A-cation ordering. The dotted-line boxes at the bottom show the charge-ordering (CS or FE) and magnetic ordering (F, AF, Fi or APAF) patterns. The arrows correspond to Fe-atoms (1) to (6) from left to right, with the valence state shown by the length of the arrow, following the correspondence for each of the two charge configurations given in the upper right, and the local moment by the direction of the arrow.}
\label{fig:s-group}
\end{figure}

Fig.~\ref{fig:s-group} shows the symmetry analysis of charge and magnetic ordered phases for a superlattice with uniformly separated La-(111) layers.
We start with the $R\bar{3}c$ (167) structure, generated from the cubic perovskite structure by oxygen octahedron rotations around [111]. A-site layering along (111) lowers the symmetry to $P\bar{3}c1$ (165). This $P\bar{3}c1$ (165) structure has a 30-atom unit cell. To distinguish Fe atoms, we assign a layer number to each Fe atom so that the Fe atom in the n$^{th}$ layer is labelled Fe(n). The non-symmorphic symmetry of the space group relates Fe(n) to Fe(n+3) and the inversion symmetry relates Fe(2) to Fe(6) and Fe(3) to Fe(5), leading to the two distinct types of Fe sites, in Wyckoff position $2b$ between two Sr-(111) layers and in Wyckoff position $4d$ between La-(111) and Sr-(111) layers, as shown in Fig.~\ref{fig:s-group} (a) by silver and black spheres, respectively. 

We enumerate reasonable charge orderings commensurate with the 30-atom unit cell, combining two nominal valence states as in the majority of charge ordered oxides\cite{S_Imada98,S_Ishihara10,S_Yamauchi14}. The reasonable nominal valence state combinations of Fe are (2$\times$Fe$^{5+}$, 4$\times$Fe$^{3+}$) and (2$\times$Fe$^{3+}$, 4$\times$Fe$^{4+}$). For the Fe$^{5+}$/Fe$^{3+}$ charge configuration, the only charge ordering pattern that preserves the symmetry is obtained by assigning Fe$^{5+}$ to the silver sites ($2b$ Wyckoff position) and Fe$^{3+}$ to the black sites ($4d$ Wyckoff position), corresponding to the left side of Fig.~\ref{fig:s-group} (b). For the symmetry-lowering charge-ordering patterns, we choose the ones with uniformly separated (111)-oriented Fe$^{5+}$ planes, to be compatible with the A-site cation ordering with uniformly separated La-(111) planes. One symmetry-lowering charge ordering pattern is obtained by assigning Fe$^{3+}$ to the silver and to half of the black sites (Fe(2) and Fe(5)) and Fe$^{5+}$ to the remaining black sites (Fe(3) and Fe(6)), which lowers the space group symmetry to non-centrosymmetric $P3c1$ (158), corresponding to the right side of Fig.~\ref{fig:s-group} (b). The $4d$ Wyckoff position of $P\bar{3}c1$ splits to $2a$ and $2c$ Wyckoff positions of the space group $P3c1$. For the Fe$^{3+}$/Fe$^{4+}$ charge configuration, we similarly construct two charge ordering patterns. 

For each charge-ordering pattern, we consider either ferromagnetic or antiferromagnetic interaction between the nearest neighbor Fe sites with four possible magnetic orderings: ferromagnetic (F), antiferromagnetic (AF), antiphase antiferromagnetic (APAF) and ferrimagnetic (Fi). In the APAF phase, the adjacent Fe in the same valence state are antialigned and in different valence states are aligned, while in the Fi phase the adjacent Fe in the same valence state are aligned and in different valence states are antialigned. The combinations for charge and magnetic ordering patterns are presented in the lower part of Fig.~\ref{fig:s-group}. The magnetic orderings can be classified by whether the local moments of Fe(n) and Fe(n+3) are in the same direction (type 3 magnetic group) or in opposite directions (type 1).  

\begin{table}[htbp]
\caption{\label{tab:low-en-str} Structural parameters of CS-APAF and FE-APAF phase from the full structural relaxation.}
\centering
\begin{ruledtabular}
\begin{tabular}{ccccc|ccccc}
   \multicolumn{5}{c|}{CS-APAF ($P\bar{3}c1$ (165))} &  \multicolumn{5}{c}{FE-APAF ($P3c1$ (158))}\\
\hline
 \multicolumn{5}{c|}{$a$=5.565{\AA}, $c$=13.458{\AA}}&  \multicolumn{5}{c}{$a$=5.556{\AA}, $c$=13.510{\AA} }\\ 
Fe & 2b & 0 & 0 & 0 & Fe & 2a &0 &0 & 0.002 \\
     & 4d & 1/3 & 2/3 & 0.168 &  & 2b &1/3 &2/3 &0.163\\
O  &6f & 0.457 & 0  & 1/4  & & 2c & 2/3 &1/3 & 0.334 \\  
    & 12g & 0.195 & 0.322 & 0.084 & O  &   6d & 0.464 & 0.006 & 0.252  \\
La &  2a &  0 & 0 & 1/4   &      & 6d & 0.200 & 0.328 &0.086 \\
Sr &  4d  & 1/3 & 2/3 & 0.419 &  &  6d & 0.800 & 0.135 &0.412 \\
    &   &            &      &               &  La  & 2a &  0 & 0 & 1/4  \\
  &   &            &      &               &  Sr &  2b &  1/3 & 2/3  & 0.417\\
  &   &            &      &               &     &  2c &2/3 &1/3  & 0.082 
\end{tabular}
\end{ruledtabular}
\end{table}

For each centrosymmetric charge and magnetic ordering pattern that we consider, we perform a full structural relaxation, as discussed in the main text. We construct the starting crystal structure with breathing distortions corresponding to the charge-ordering pattern by decreasing the distance from the higher-valence-Fe plane to the adjacent O planes by 0.04 {\AA}. We set the starting magnetic ordering pattern with the magnitudes of the magnetic moments equal to that of the nominal valence high-spin state. We obtain five distinct locally stable phases, labelled CS in Table 1. For Fe$^{5+}$/Fe$^{3+}$ charge configuration, we find locally stable phases for all the magnetic ordering patterns except the ferromagnetic ordering whereas for Fe$^{4+}$/Fe$^{3+}$ charge configuration we find only one locally stable phase with ferrimagnetic ordering in which initial structures with APAF and AF ordering relax to Fe$^{5+}$/Fe$^{3+}$ charge configuration maintaining the same magnetic ordering pattern. We note that with ferromagnetic ordering, for both charge configurations the starting structure relaxes to the CS-F phase with all Fe having same charge state (Fe$^{+3.67}$). For the ferroelectric charge ordering patterns, we consider the eight charge and magnetic ordering patterns from Fig.~\ref{fig:s-group}. If the pattern corresponds to one of the five locally stable CS phases, we generate the starting structure by taking the relaxed CS structure and interchanging the A-site cations to shift the registry between the cation ordering and the charge ordering. If not, we generate the starting crystal structure with breathing distortions as described above. In this case, full structural relaxation yields three distinct locally stable phases, labelled FE in Table 1. For ferromagnetic ordering, both charge configurations relax to CS-F phase discussed above. For Fe$^{5+}$/Fe$^{3+}$ charge configuration, we find the locally stable phases with APAF and AF orderings whereas the CS-Fi phase relaxes to FE-Fi2 phase. For Fe$^{4+}$/Fe$^{3+}$ charge configuration, we find no locally stable phase in which the APAF and AF phases relax to Fe$^{5+}$/Fe$^{3+}$ charge configuration maintaining the same magnetic ordering pattern and the Fi phase relaxes to FE-Fi2 phase with Fe$^{5+}$/Fe$^{3+}$ charge configuration while maintaining the same total magnetization of 6 $\mu_{B}$ of the 30-atom unit cell.  
Table \ref{tab:low-en-str} shows the computed structural parameters of the CS-APAF and FE-APAF phases.

\section{Symmetry-protected ferrimagnetic metallic phase}

\begin{figure}[htbp]
\begin{center}
\includegraphics[width=0.8\columnwidth, angle=-0]{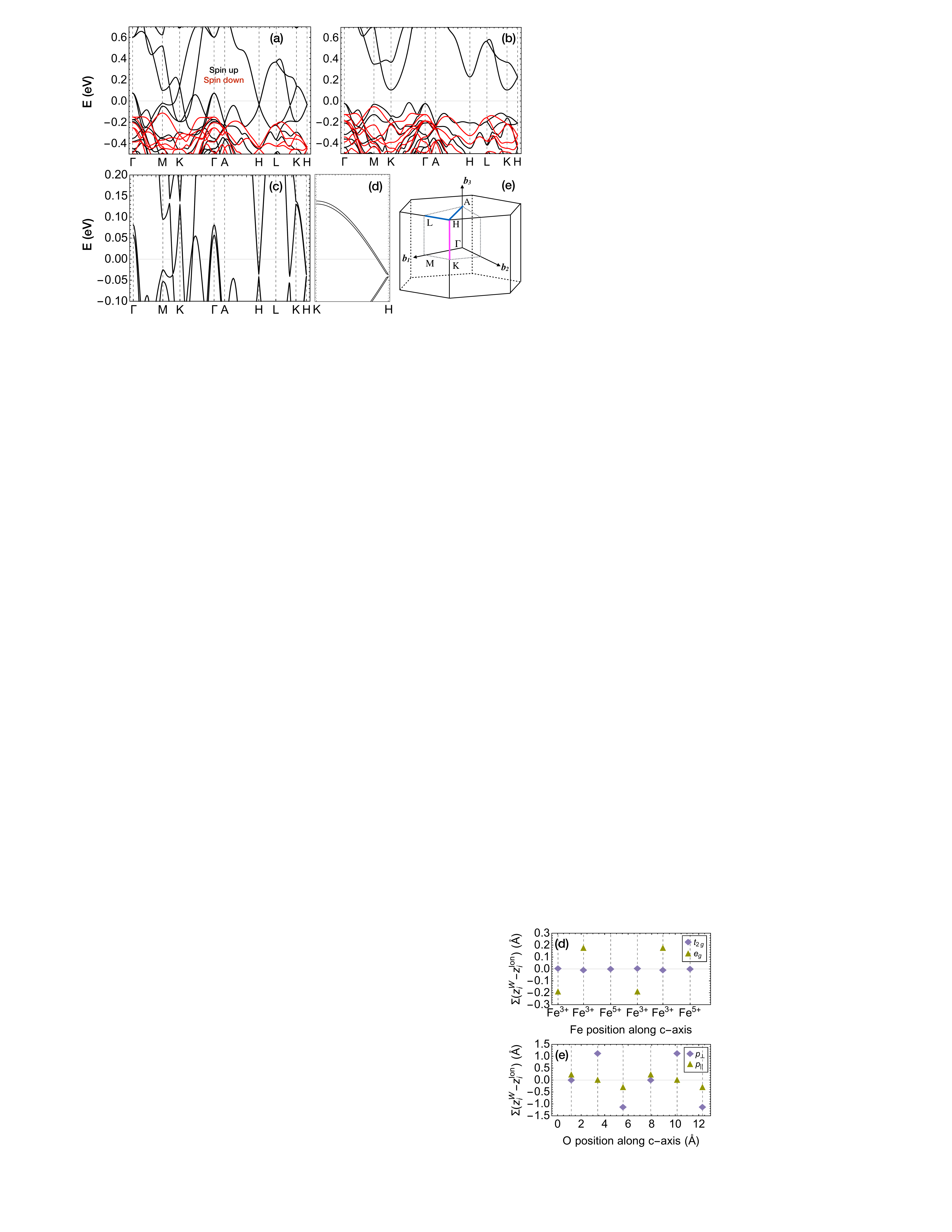}
\caption{Band structures along the high-symmetry lines for (a) CS-Fi phase with $P\bar{3}c^{\prime}1$ space group (2$\times$Fe$^{3+}$/$4\times$Fe$^{4+}$) and (b) FE-Fi2 phase with $P3c^{\prime}1$ space group (2$\times$Fe$^{5+}$/$4\times$Fe$^{3+}$) calculated without spin-orbit coupling. The black and red lines are spin up and spin down bands, respectively. (c) Band structures of CS-Fi phase around the Fermi energy with spin-orbit coupling. (d) Expanded view of he band splitting along the K-H line shown in (c). (e) Brillouin zone of the hexagonal lattice. The solid blue and magenta  lines are $k$-lines for which the energy eigenvalues in CS-Fi phases are doubly degenerate. Breaking of the inversion symmetry by transition to the FE-Fi2 phase and/or inclusion of spin-orbit coupling lifts the double degeneracy along the magenta line, while the double degeneracy along the blue line is maintained.}
\label{fig:bd-fi}
\end{center}
\end{figure}

The computed band structure of the centrosymmetric ferrimagnetic phase (CS-Fi, see last column of Table I)  with
2$\times$Fe$^{3+}$/$4\times$Fe$^{4+}$ charge ordering and space group $P\bar{3}c^\prime1$ (type 3 magnetic group)
is shown in Fig.~\ref{fig:bd-fi} (a). The system is half metallic, with only the majority spin states crossing the Fermi energy.
Without spin-orbit coupling, we note a 4-fold degenerate Dirac-like point at H and
doubly-degenerate bands crossing the Fermi energy along the A-H, H-L, and K-H lines as shown in Fig.~\ref{fig:bd-fi} (a). The double degeneracy along the K-H line marked with a magenta line in the Brillouin zone is protected by three-fold rotation symmetry around the $c$-axis, inversion, and time reversal (TR) symmetry, while the double degeneracy along the A-H (H-L) lines, lying in the zone boundary plane marked with the blue line in the Brillouin zone, is protected by TR combined with $c$-axis glide with two-fold rotation along the [110] direction ($c$-axis glide with (110) mirror). When the inversion symmetry is broken by interchanging a pair of La and Sr planes, this phase relaxes to the 2$\times$Fe$^{5+}$/$4\times$Fe$^{3+}$ FE-Fi2 phase (see Table I), lowering the total energy about 9 meV/Fe and opening the band gap, as shown in Fig.~\ref{fig:bd-fi} (b). The resulting space group is $P3c^{\prime}$1 (type 3 magnetic group). Breaking the inversion symmetry lifts the double degeneracy of the K-H line crossing the Fermi energy, while the degeneracies along the A-H and H-L lines are preserved.
Including spin-orbit coupling aligns the magnetization along the $c$ direction with a magnetocrystalline anisotropy energy (MAE) less than 20 $\mu$eV relative to magnetization in the $ab$ plane (MAE within the $ab$ plane is an order smaller), preserving the combination of TR and glide-rotation (glide-mirror) symmetries of the magnetic space group. In both phases, the double degeneracy along the A-H and H-L line is maintained, but the degeneracy along the K-H line in the CS phase is lifted, and there is splitting of the 4-fold degeneracies at H and K of the order of 10 meV as shown in Fig.~\ref{fig:bd-fi} (c-d).
Similar features in the band structure are also observed in (LaXO$_3$)$_2$/(LaAlO$_3$)$_4$(111)-bilayer superlattices with ferromagnetically-coupled  X-$d^{4}$ bilayers\cite{S_Doennig16}.

\section{Wannier center analysis of polarization}
We can understand the nature of the switching polarization more intuitively by using the Wannier function formulation for the electronic contribution to the polarization\cite{S_King-Smith93}
\begin{eqnarray}
\mathbf{P}_{el} = -\frac{e}{\Omega}  \sum_{n} \int d^{3}r \; \mathbf{r} \vert W_{n}(\mathbf{r}) \vert^{2}= -\frac{e}{\Omega} \sum_{n} \mathbf{R_{n}}
\end{eqnarray}
where $\Omega$ is the unit cell volume, $W_{n}(\mathbf{r})$ is the n$^{th}$ Wannier function, and $\mathbf{R_{n}}$ denotes the center of charge of the Wannier function (Wannier center).

We used wannier90\cite{S_Mostofi14} to obtain Wannier functions and centers for the valence bands including Fe-$d$ and O-$p$ bands, with a total of 134 occupied bands (6 electrons for each of 18 oxygens, 5 electrons for four Fe$^{3+}$ ions, and 3 electrons for two Fe$^{5+}$ ions). Wannier centers for bands at lower energy (60 bands (120 electrons), consisting of Fe[$3p^{6}$], O[$2s^{2}$], La[$5s^{2}5p^{6}$], and Sr[$4s^{2}4p^{6}$]) are assumed to be located at the atomic positions.

\begin{figure}[htbp]
\begin{center}
\includegraphics[width=0.7\columnwidth, angle=-0]{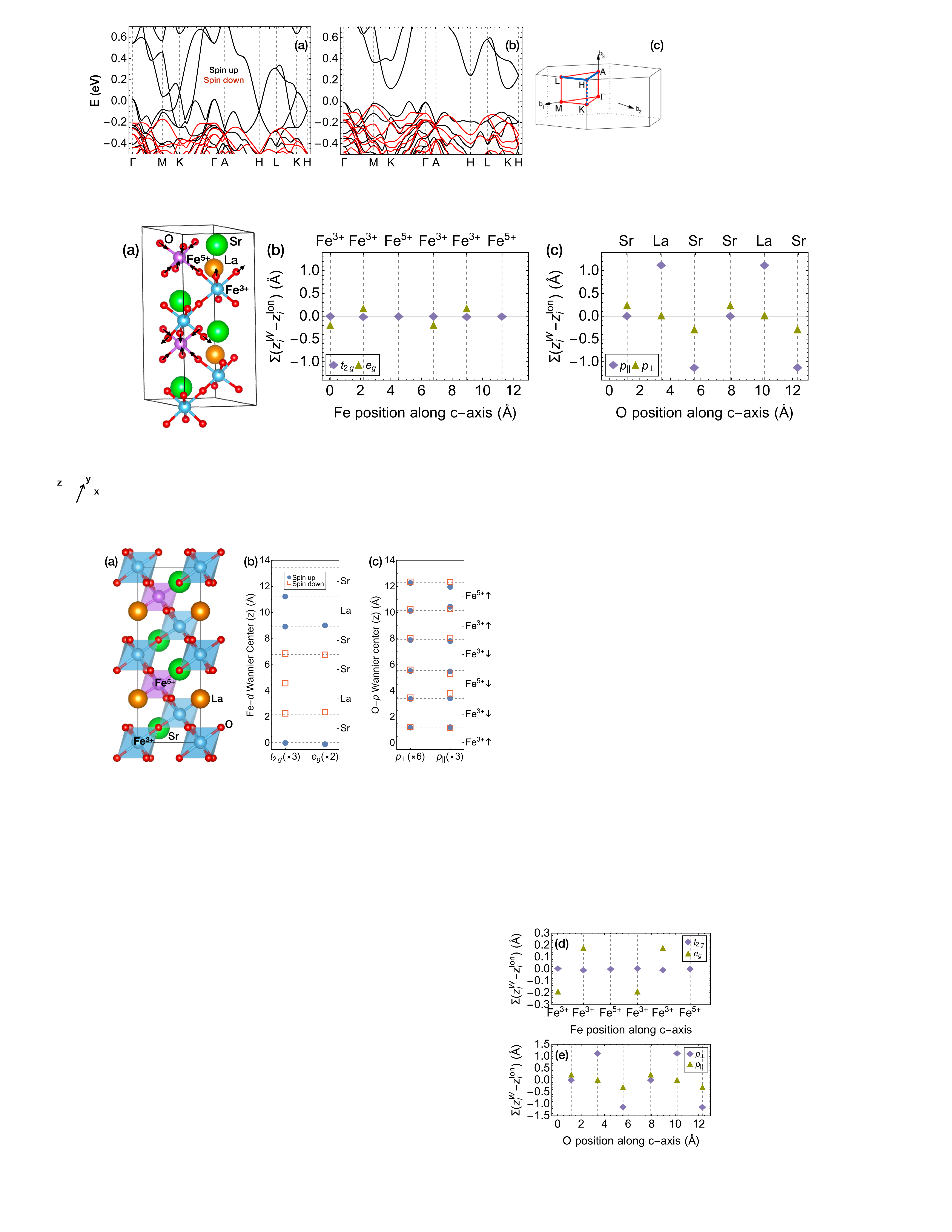}
\caption{ (a) Side view of the atomic arrangement of the FE-APAF phase. (b) Average $z$-coordinates of occupied Fe-$t_{2g}$ and Fe-$e_{g}$ Wannier centers in {\AA}. Each datapoint for $t_{2g}$ and $e_{2g}$ orbitals is averaged over three $t_{2g}$ and two $e_{g}$ orbitals defined with respect to the local octahedral axes. Dashed lines denote the $z$ coordinates of Fe atoms. The $z$-coordinates of La and Sr cations are shown in the right vertical axis. (c) $z$-coordinates of the O-$p$ Wannier centers, averaged in the relevant plane. Each datapoint represents the $z$-coordinates of Wannier centers, averaged over six O-$p_{||}$ and three $p_{\perp}$ orbitals for each spin channel, respectively. Dashed lines denote the $z$ coordinates of O atoms. The symbols $p_{||}$ and  $p_{\perp}$ denote O-$p$ orbitals along and perpendicular to the Fe-O bonding direction, respectively.  The $z$-coordinates of Fe atoms are shown in the right vertical axis with nominal valence and spin direction.}
\label{fig:wfcenters}
\end{center}
\end{figure}

Fig.~\ref{fig:wfcenters} shows the $z$ coordinates of O-$p$ and Fe-$d$ Wannier centers. As expected, there are six O-$p$ Wannier functions per oxygen site. For Fe-$d$, we find that there are 5 and 3 Wannier functions around the Fe$^{3+}$ and Fe$^{5+}$ sites, respectively, corresponding to $d^{5}$/$d^{3}$ charge ordering. 
This allows us to understand the polarization switching as a transfer of two electrons from an Fe$^{3+}$ site to the neighboring Fe$^{5+}$ site as in Fig. 5.  With two such transfers per unit cell, we can then estimate the switching polarization by multiplying the total transferred charge of 4e by one sixth of the $c$ lattice constant, resulting in 40 $\mu$C/cm$^{2}$, and thus largely accounting for the switching polarization of 39 $\mu$C/cm$^{2}$ computed by the Berry phase method.  

\begin{figure}[htbp]
\begin{center}
\includegraphics[width=0.95\columnwidth, angle=-0]{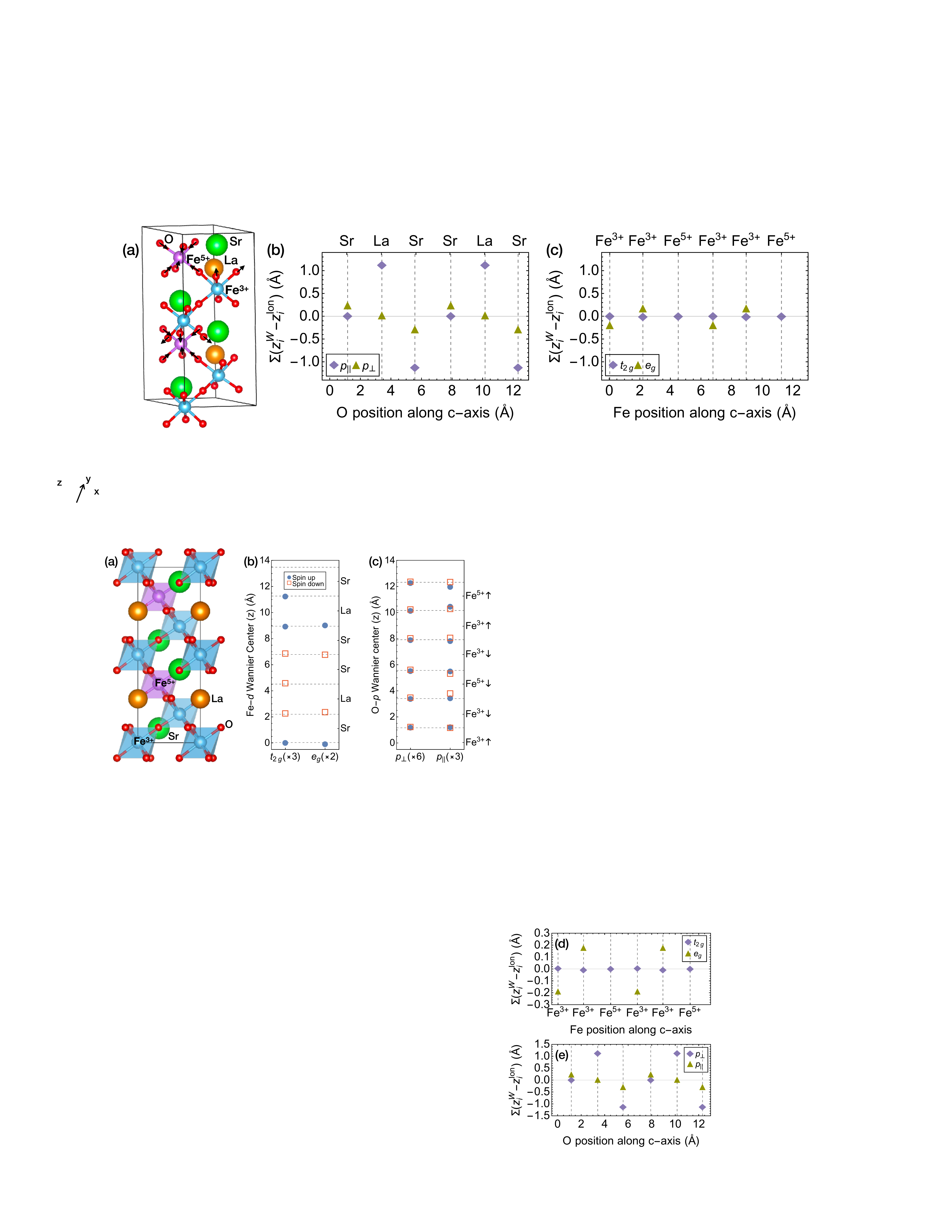}
\caption{(a) Illustration of the shifts of O-$p_{||}$ Wannier centers from oxygen atomic positions,  represented by arrows. Displacements of Wannier centers from the atomic positions along the $c$-axis summed over O-$p$ Wannier centers (panel (b)) and  Fe-$d$ Wannier centers (panel (c)) in the relevant (111)-plane. For Fe-$d$ orbitals, the blue diamond and green triangle symbols represent 3 $t_{2g}$ and 2 $e_{g}$ Wannier centers, respectively. Similarly for O-$p$ orbitals, the blue diamond and green triangle symbols represent 6 $p_{||}$ and 12 $p_{\perp}$ Wannier centers, respectively. }
\label{fig:wann}
\end{center}
\end{figure}

As shown in Fig.~\ref{fig:wfcenters} (c) and Fig.~\ref{fig:wann} (a,b), there are substantial shifts in the O-$p$ Wannier functions lying along the Fe-O bonding direction denoted by $p_{||}$. 
The shift of O-$p_{||}$ Wannier centers by up to 0.4 {\AA}  into the Fe$^{5+}$ sites is due to the screening by O-$p_{||}$ orbitals, consistent with a picture of oxygen ligand holes with strong mixing between empty Fe$^{5+}$-$e_{g}$ and filled O-$p_{||}$ states. 
In Fig.~\ref{fig:wfcenters} (b) and Fig.~\ref{fig:wann} (c), we plot the positions and displacements from atomic positions for Fe-$d$ Wannier centers. The Wannier centers with $t_{2g}$ symmetry are located at the atomic positions, while the centers with $e_{g}$ symmetry on Fe$^{3+}$ are shifted toward the Fe$^{5+}$ plane. This shift is due to the strong hybridization between O-$p_{||}$ and Fe$^{3+}$-$e_{g}$ orbitals.
 Finally, due to the relatively weak hybridization between O-$p_{\perp}$ and Fe$^{3+}$-$e_{g}$ orbitals, O-$p_{\perp}$ Wannier centers show relatively quite small shifts. We note that the contribution of the Wannier center shifts to the switching polarization, like the contributions of ionic displacements, are quite small relative to the contribution of the inter-Fe electron transfer discussed above.

\section{Dependence of the total energies of low energy phases on the on-site Coulomb interaction}
Our computations were performed with a value of $U$ = 5.4 eV for Fe $d$ orbitals. 
Table \ref{tab:en-vs-u} presents the total energies computed with $U$ values for Fe ranging from 4.4 eV to 6.4 eV. We find that throughout this range, the energy difference between the CS-APAF and FE-APAF phases is less than 3 meV, showing that the small energy difference between the centrosymmetric and ferroelectric charge ordering patterns is robust against changes in the precise value of $U$. We note that the energy of the metallic CS-F phase becomes close to the CS-APAF phase for $U$ = 6.4 eV, possibly due to the energy gain from the increased exchange splitting. 

\begin{table}[htbp]
\caption{\label{tab:en-vs-u} Total energy per Fe (in meV) for $U$ values ranging from 4.4 to 6.4 eV. Values are reported relative to the lowest energy state at each $U$. The value of $J$ is fixed to 1 eV. The charge and magnetic ordering patterns which relax to other low energy phases are denoted as unstable.}
\centering
\begin{ruledtabular}
\begin{tabular}{c||cccccc|c|c}
$U$ (eV) &\multicolumn{6}{c|}{2Fe$^{5+}(d^{3})$/4Fe$^{3+}$($d^{5}$)}  & 6Fe$^{3.67+}$($d^{4.33}$)  & 2Fe$^{3+}$($d^{5}$)/4Fe$^{4+}$($d^{4}$) \\
\cline{2-9}
 &CS-APAF & FE-APAF & CS-AF & FE-AF & CS-Fi  & FE-Fi & CS-F & CS-Fi  \\
\hline
 4.4 & 1.8 & 0 & 108 & 103 & 240 & 30 & 37 & 39 \\
 4.9 & 0.5 & 0 & 111 & 107 & 237 & 32 & 26 & 41 \\
 5.4 &  0 & 0.5 & 124 & 119 & 243 & 42 & 24  & 51 \\ 
 5.9 &  0 & 1.5 & 124 & 118 & unstable & 38 & 8.4 & 45\\
 6.4 &  0 & 2.3 & 127 & 122 & unstable & 38 & 1.5 & 45\\
\end{tabular}
\end{ruledtabular}
\end{table}

\end{document}